\documentclass[12pt]{article}
\usepackage{amsmath}
\usepackage{amssymb}
\usepackage{graphicx}
\newcommand{\deriv}[2]{ \frac{d #1}{d #2} }

\newcommand{\mb}[1]{\boldsymbol{#1}}
\newcommand{\arcch}{\cosh^{-1}}

\title{The Gravitational Cherenkov Radiation}
\author{A.M. Ignatov
\thanks{General Physics Institute, 38 Vavilova str., Moscow, Russia.
E-mail: aign@fpl.gpi.ru.
 }}
\begin{document}

\maketitle

\begin{abstract}
An example of discontinuity of the energy-momentum tensor moving
at superluminal velocity is discussed. It is shown that the
gravitational Mach cone is formed. The power spectrum of the
corresponding Cherenkov radiation is evaluated.
\end{abstract}

\section{Introduction}
According to current beliefs the traditional relativity theory
excludes the possibility of superluminal motion.  More detailed
analysis shows that this constraint actually concerns either
information exchange between various events in a space-time or
motion of material bodies. It should be noted however that,
strictly speaking, the concept of information is beyond physics as
long as there is no physical definition (except purely
tautological ones)  of a material body. Hyperbolizing a bit one
may say that relativity imposes no limitations on  propagation
velocity of physical quantities like charge, mass \textit{etc.}

An everyday life  example of a superluminal charge motion, which
may be found in many physical laboratories, is an image of an
electron beam at an oscilloscope screen. With an oscilloscope
operating, say, at 1 GHz sweep frequency, the velocity of the
charge spot would exceed the speed of light if the screen is more
than 30 cm wide. Another example is a light spot scanning across a
dielectric surface and producing, therefore, the superluminal
polarization wave. Evidently, these are the examples of
superluminal \textit{phase} velocities and there are no
contradictions with the conformist approach to causality. However,
from the viewpoint of electrodynamics a  charge spot behaves like
a real superluminal charge and it may emit the Cherenkov radiation
even in vacuum.

Seemingly, for the first time the Cherenkov radiation of a
reflected light spot was discussed by Franck \cite{franck}. Later,
a number of detailed theoretical studies was performed; see,
\textit{e.g.,} the review papers \cite{bg,ginzburg,bars}. The
rumor runs that  this radiation was experimentally observed;
however, I failed to find the reference.

It was pointed out (\textit{e.g.,}\cite{bg}) that similar
gravitational radiation may be emitted by a gravitational wave
spot or a superluminal mass spot. However, to my knowledge, there
was no detailed study of the emitted gravitational wake. In the
present note the linearized Einstein equations are used to
investigate the structure of  gravitational field behind a
superluminal mass spot.

\section{Model}

A model discussed in this paper is depicted in Fig.~1. Imagine two
infinitesimally thin massive threads moving  at an angle towards
the \textit{x}-axis. The threads collide at a certain point and
form a new composite thread due to some inelastic process,
\textit{e.g.,} a chemical reaction. At an instant of collision
every atom experiences infinite deceleration that, presumably,
should give rise to the gravitational bremsstrahlung.

Let's choose the reference frame so that the velocities of the
 threads are  $\mb{\beta}=\beta \mb{n}=\beta(\cos\theta,
\sin\theta,0)$  and $\mb{\beta}^\prime=\beta
\mb{n}^\prime=\beta(\cos\theta, -\sin\theta,0)$.  The mass
densities of both threads are equal and in the chosen frame the
net density  prior to the collision is given by $j^0(\xi)=\sigma
\delta(z)[ \delta(\mb{n} \cdot \mb{r}-\beta
t)+\delta(\mb{n}^\prime \cdot \mb{r}-\beta t)]$, where $\sigma$ is
the mass per unit length
 and $\xi=(t, x,y, z)=(t,\mb{r})$. It is assumed that $0<\theta<\pi/2$
 and the speed of light is unity. The flat space-time
 metric is used, $ds^2=\eta_{\mu\nu}d\xi^{\mu}d\xi^{\nu}=-dt^2+d\mb{r}^2$.

\begin{figure}
\centerline{\includegraphics{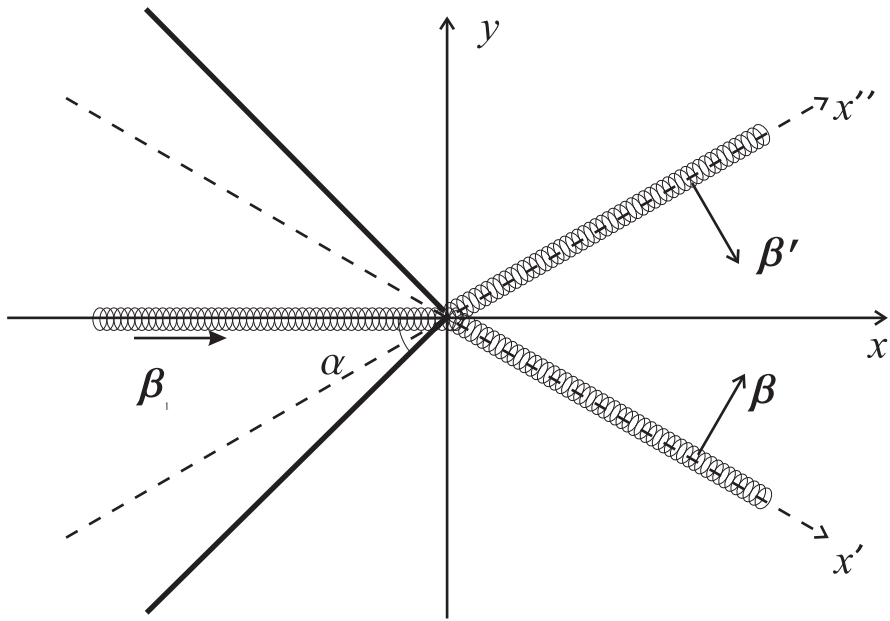}} \caption{Two colliding
massive threads depicted as spirals produce the gravitational
radiation at the Mach cone (heavy line). \label{fig1}}
\end{figure}

  Let $x_0(t)$ be the coordinate of the
intersection point; evidently, $x_0(t)=\beta t/ \cos\theta \equiv
u t$. Depending on the angle of incidence, $\theta$, the velocity,
$u$, of the intersection point may be either subluminal or
superluminal.

 Let the threads be composed of a zero-pressure ideal gas.
Therefore,  the energy-momentum tensor prior to the collision
  is written as
\begin{equation}\label{t+}
T^{(+)\mu\nu}(\xi)=\sigma\delta(z)\left[ \delta(\mb{n} \cdot
\mb{r}-\beta t)\tau^{\mu\nu}(\mb{\beta})+\delta(\mb{n}^\prime
\cdot \mb{r}-\beta t)\tau^{\mu\nu}(\mb{\beta}^\prime)\right],
\end{equation}
where
 $$ \tau^{00}(\beta)=\gamma(\beta),\qquad
\tau^{0i}(\beta)=\gamma(\beta)\beta^i, \qquad
\tau^{ij}(\beta)=\gamma(\beta) \beta^{i}\beta^j,$$ and
$\gamma(\beta)=1/\sqrt{1-\beta^2}$.

Due to the symmetry of the problem the composite thread formed
after the collision is situated at the \textit{x}-axis and its
energy-momentum tensor may be written as

\begin{equation}\label{t-}
T^{(-)\mu\nu}(\xi)=\sigma_1\delta(z)\delta(y)\tau^{\mu\nu}(\mb{\beta}_1).
\end{equation}
 Integrating  (\ref{t+},\ref{t-}) over the appropriate
  two-dimensional surfaces we find that the requirement of
   energy-momentum conservation
  results in
\begin{equation}\label{s1}
\mb{\beta}_1=(\beta \cos\theta, 0,0),\qquad \sigma_1=\frac{2\sigma
\gamma(\beta) }{\sin\theta \gamma(\beta_1)},
\end{equation}
that is, the \textit{x}-component of the thread velocity remains
unchanged.
   It should be pointed out that we expect that a
part of the kinetic energy of the colliding threads is carried
away by the gravitational bremsstrahlung. However, in the linear
approximation, which we are going to implement, the influence of
radiation losses on the motion and the mass of the composite
thread is evidently negligible.

Thus, the net energy-momentum tensor is

\begin{equation}\label{enmom}
 T^{\mu\nu}(\xi)=T^{(+)\mu\nu}(\xi)\vartheta(x-u t)
 +T^{(-)\mu\nu}(\xi)\vartheta(u t -x),
\end{equation}
where $\vartheta(x)$ is the Heaviside's step function.

It should also be noted that although this model provides
conservation of energy and momentum, $T^{\mu\nu}_{\ \ ,\,\nu}=0$,
the mass  is a non-conserving quantity: $\sigma_1>2\sigma$ in
Eq.~(\ref{s1}). This is the evident consequence of including
non-gravitational interactions (\textit{e.g.,} chemical reactions)
into the model that results in the mass defect. On the other hand,
we can easily think out various  models with mass conservation and
energy losses. This may be, for example, a single massive thread
moving at an angle towards an immobile absolutely rigid adhesive
screen. The most essential feature of these models is the
discontinuity of the energy-momentum moving at superluminal phase
velocity. The problem is what kind of gravitational field is
produced by the discontinuity.

\section{Mach cone}

 Our goal here is to analyze the solution of the linearized Einstein
 equations with the energy-momentum tensor given by
 Eq.~(\ref{enmom}).
 Keeping the notation of \cite{mtw} the d'Alembert equation for
 the distortion of the metric tensor is
 \begin{equation}
 -\bar{h}^{\mu\nu}_{\ \  ,}{}^{\ \lambda}_{ \lambda}=16 \pi
 T^{\mu\nu},
 \label{dAlembert}
 \end{equation}
where
\begin{equation}\label{h}
g^{\mu\nu}=\eta^{\mu\nu}+h^{\mu\nu}\equiv\eta^{\mu\nu}+\bar{h}^{\mu\nu}-\frac12
\eta^{\mu\nu} \bar{h}^{\lambda}_{\ \lambda}.
\end{equation}

Since the energy-momentum tensor (\ref{enmom}) is composed of
three parts, the solution of Eq.~(\ref{dAlembert}) is also a
superposition of three retarded solutions. First, consider the
solution of the d'Alembert equation conditioned by the composite
thread at $x<u t$. Let $f^{(-)}(\xi) $ be the retarded solution of
the equation

\begin{equation}\label{fe-}
\square f^{(-)}(\xi)=4 \pi \vartheta(u t-x)\delta(y)\delta(z)
\end{equation}
that explicitly is written as
\begin{equation}
f^{(-)}(\xi)= \int\limits^\infty_{-\infty} \frac{d
s}{\sqrt{s^2+\rho^2}} \vartheta\left(-s-u \sqrt{s^2+\rho^2}+u \tau
\right),\label{f-}
\end{equation}
where $\tau=t-x/u$ and $\rho=\sqrt{y^2+z^2}$. The integral in
Eq.~(\ref{f-}) essentially depends on the velocity of the tip of
the thread, $u$. In what follows we focus on the most interesting
superluminal motion. If $u>1$ then
\begin{equation}
f^{(-)}(\xi)=2 \vartheta(\tau -\cos\alpha\rho)
\arcch\left(\frac{\tau}{\cos\alpha \rho}\right),\label{f-+}
\end{equation}
where $\sin\alpha=1/u$, \textit{i.e.} the result is non-zero
inside the Mach cone
\begin{equation}\label{mach}
\tau\geq\cos\alpha\rho
\end{equation}
depicted by the heavy line in Fig.~1.

In order to evaluate the field produced by the colliding threads
at $x>ut$ note that

$$\vartheta(x-u t)\delta(\mb{n} \cdot \mb{r}-\beta
t)=\gamma(\beta)\delta(y^\prime)\vartheta(x^\prime-u^\prime
t^\prime),$$ where $t^\prime,x^\prime,y^\prime,z^\prime$ are own
frame coordinates of the moving thread. The corresponding Lorentz
transform is composed of a boost and a rotation around the $z$
axis:
\begin{eqnarray}
t^\prime&=&\gamma(t-\beta \mb{n} \cdot \mb{r})\nonumber\\
x^\prime&=& x \sin\theta-y\cos\theta\nonumber\\
y^\prime&=&\gamma(-\beta t +\mb{n} \cdot \mb{r})\label{lt}\\
z^\prime&=&z\nonumber
\end{eqnarray}

 The speed of the tip of the moving part of the thread in its own
 frame is $u^\prime=\beta\gamma\tan\theta$; evidently, if $u>1$
 then $u^\prime>1$ and vice versa.

 The solution of the d'Alembert equation
 \begin{equation}\label{fe+}
\square f^{(+)}(\xi)=4 \pi \vartheta(x-u t)\delta(\mb{n} \cdot
\mb{r}-\beta t)\delta(z)
\end{equation}
is  given by
\begin{eqnarray}
f^{(+)}(\xi)&=& \gamma(\beta) \int\limits^\infty_{-\infty} \frac{d
s}{\sqrt{s^2+\rho^{\prime 2}}} \vartheta\left(s+u^\prime
\sqrt{s^2+\rho^{\prime2}}-u\tau^\prime\right)\label{f+}\\
 &=&\gamma(\beta)\left[-2 \ln \rho^\prime-2 \vartheta(\tau^\prime- \cos\alpha^\prime \rho^\prime)
\arcch\left(\frac{\tau^\prime}{\cos\alpha^\prime\rho^\prime}\right)\right],
\label{f++}
\end{eqnarray}
where $\tau^\prime=t^\prime-x^\prime/u^\prime$,
$\rho^\prime=\sqrt{y^{\prime 2}+z^2}$ and
$\sin\alpha^\prime=1/u^\prime$.  One can easily verify that
inequalities $\tau^\prime\geq\cos\alpha^\prime\rho^\prime$ and
(\ref{mach}) defines the same conical area in space.

 Comparing
Eqs.~(\ref{f-},\ref{f+}), we see that due to the different signs
of the arguments of the $\vartheta$-functions the integral
(\ref{f+}) is a composition of the logarithmic potential produced
by the infinitely thin massive thread and the  wake wave analogous
to the one excited by the composite part (\ref{f-+}). However
unlike Eq.~(\ref{f-+}), the structure of the wake field
(\ref{f++}) is not axially-symmetric in the initial reference
frame. It should also be noted that there is no logarithmic
singularity at $x^\prime<0$ in Eq.~(\ref{f++}), i.e. at the left
side of the  $x^\prime$-axis shown by the dashed line in Fig.~1,
which is always inside the Mach cone.

The field of the second thread is evaluated in the same way.
Finally, the resulting metric perturbation is given by

\begin{eqnarray}
\bar{h}^{\mu\nu}(\xi)&=&-8\sigma\gamma(\beta)\left\{
\tau^{\mu\nu}(\mb{\beta}) \left[ \ln \rho^\prime+\vartheta(\tau-
\cos\alpha \rho)
\arcch\left(\frac{\tau^\prime}{\cos\alpha^\prime\rho^\prime}\right)\right]
\right.\nonumber\\
&+& \left. \tau^{\mu\nu}(\mb{\beta}^\prime) \left[   \ln
\rho^{\prime\prime}+ \vartheta(\tau- \cos\alpha \rho)
\arcch\left(\frac{\tau^{\prime\prime}}{\cos\alpha^\prime\rho^{\prime\prime}}
\right)\right] \right\} \nonumber \\
&+&\frac{16 \sigma\gamma(\beta)}{\sin\theta \, \gamma(\beta_1)}
\tau^{\mu\nu}(\mb{\beta}_1) \vartheta(\tau -\cos\alpha\rho)
\arcch\left(\frac{\tau}{\cos\alpha \rho}\right),\label{sol}
\end{eqnarray}
where the double primed variables are given by Eqs.~(\ref{lt})
with $\theta$ replaced by $-\theta$.

Although Eq.~(\ref{sol}) looks pretty bulky its structure is
transparent. The field ahead of the collision point, $x>ut$ is a
superposition of two  logarithmic potentials distorted  by the
Lorentz transform. The field inside the Mach cone  (\ref{mach})
also looks like the static one near the $x$-axis. However, the
metric and some of the components of the Riemann tensor are
divergent at the Mach cone. This is the evident consequence of the
accepted model: taking into account the finite size of the threads
would smooth this singularity.

\section{Radiation field}

The discussed collision process produce the Mach cone at the
space-time fabric. However, it is obtained in the near zone and it
is unclear whether there is any radiation field. The
straightforward evaluation of the gravitational energy-momentum
flux using Eq.~(\ref{sol}) yields very complicated integrals. We
can circumvent  this difficulty implementing the procedure similar
to the one used in electrodynamics \cite{bg,ginzburg}.

Suppose that the  system as a whole is of large but finite size,
$L$. Performing  the time Fourier transform of
Eq.~(\ref{dAlembert}) the asymptotic of $\bar{h}^{\mu\nu}(\omega,
\mb{r} )$ is written as ( \textit{e.g.,} \cite{ll})

\begin{equation}
\left. \bar{h}^{\mu\nu}(\omega, \mb{r} )\right|_{r \gg L}=
\frac{e^{i \omega r}}{r} \bar{\Lambda}^{\mu\nu}(\omega,
\mb{s}),\label{as}
\end{equation}
where $\mb{s}=\mb{r}/r$ is the direction of the wave propagation
and $ \Lambda^{\mu\nu}(\omega, \mb{s})$ is the Fourier transform
of the energy-momentum tensor (\ref{enmom}),

\begin{eqnarray}\label{Lambda}
& & \bar{\Lambda}^{\mu\nu}(\omega, \mb{s})=4 \int dt\,d\mb{r}\,
e^{i\omega
 t-i \omega \mb{s r}}\, T^{\mu\nu}(t,\mb{r})\\
&=&\frac{8 \pi i \sigma}{\omega^2} \delta(s_x-1/u) \left\{
\frac{\tau^{\mu\nu}(\mb{\beta})}{s_y -\sin\theta}+
\frac{\tau^{\mu\nu}(\mb{\beta}^\prime)}{s_y +\sin\theta}+\frac{2
\gamma(\beta)}{\sin\theta\, \gamma(\beta_1)}
\tau^{\mu\nu}(\mb{\beta}_1)\right\}.\nonumber
 \end{eqnarray}

Extracting the transverse traceless part of the metric
perturbation, $h^{\mu\nu}$ (\ref{h}), we obtain its spatial
components
\begin{equation}
h^{TT}_{ij}=\frac{e^{i\omega t}}{r} \frac{8\pi i
\sigma}{\omega^2}\delta(s_x-1/u)Q_{ij},
\end{equation}
where
\begin{equation}
Q_{ij}= \frac{P_{ij}(\mb{\beta})}{s_y -\sin\theta}+
\frac{P_{ij}(\mb{\beta}^\prime)}{s_y +\sin\theta}+\frac{2
\gamma(\beta)}{\sin\theta\, \gamma(\beta_1)} P_{ij}(\mb{\beta}_1)
\end{equation}
and
\begin{equation}
P_{ij}(\mb{\beta})=\frac12 \left(\delta_{ij}-s_i
s_j\right)+\gamma(\beta) (\beta_i-s_i) (\beta_j-s_j) .
\end{equation}

Since the energy-momentum of the gravitational waves is
$T^{GW}_{\mu\nu}=\langle h^{TT}_{ij,\mu} h^{TT}_{ij,\nu}\rangle /
32\pi$, evaluating its $tr$-component and
 integrating over a sufficiently large sphere we get
the power spectrum of the emitted radiation
\begin{equation}\label{power}
\deriv{W_{\omega}(\chi,\psi)}t = \frac{2 u \sigma^2}{\omega}
\delta(\cos\chi-1/u) Q_{ij}Q_{ij} \sin\chi d\chi\, d\psi\,
d\omega,
\end{equation}
where $\chi , \psi $ are spherical angles corresponding to the
direction of the wave propagation, $\mb{s}=(\cos\chi,\;
\sin\chi\,\cos\psi, \; \sin\chi\,\sin\psi)$.

The power spectrum (\ref{power}) is of typical Cherenkov
character. The $\delta$-function indicates that the  waves obeying
the resonance condition $\omega=k_x u$ only are emitted. The
complicated angular dependence hidden in $Q_{ij}^2$ is specific
for the accepted model.

Due to the $1/\omega$ dependence in Eq.(\ref{power}) the net
emitted power diverges. (The same problem also arises in
electrodynamics.) The effective cut-off may be provided by finite
length of the threads that makes integrals convergent in the
low-frequency band. The high-frequency cut-off arises due to the
finiteness of the deceleration in the process collision and/or due
to the finite diameter of the threads.

\section{Conclusion}
We have shown that a superluminal discontinuity of the
energy-momentum tensor results in the formation of the
gravitational Mach cone and the corresponding gravitational
Cherenkov radiation. Although we took special care of the energy
conservation, the problem is not self-consistent: a part of the
energy and momentum is carried away by gravitational radiation.
This would result  in a reaction force acting upon the emitting
system.

\end{document}